\documentclass[12pt]{article}
\usepackage{geometry}                
\usepackage{setspace}

\usepackage{graphicx}
\usepackage{amsmath} 
\usepackage{amssymb}
\usepackage{array}

\usepackage{color}

\def\C{{\cal C}}

\usepackage{hyperref}

\title{
Relativistic Causality and Clockless Circuits 
\footnote{\copyright \ ACM, 2011. This is the author's version of the work. It is posted here by permission of ACM for your personnal use. Not for redistribution. The definitive version was published in ACM Journal of Emerging Technologies in Computing Systems, Vol. 7, No. 4, Article 20, December 2011, \url{http://doi.acm.org/10.1145/2043643.2043650}.}
}

\author{Philippe~Matherat\\
{CNRS-LTCI (UMR 5141) and Institut T\'el\'ecom}\\
\url{http://matherat.net/}\\
\\
Marc-Thierry~Jaekel\\
{CNRS-LPTENS (UMR 8549), Ecole Normale Sup\'erieure}\\
{and Universit\'e Pierre et Marie Curie}
}

\date{}

\begin{document}

\maketitle

\begin{abstract}
Time plays a crucial role in the performance of computing systems. 
{The accurate modelling}  of logical devices, and of their physical implementations, requires
 an appropriate representation of time and of all properties that depend on this notion. 
{The need for a proper model}, particularly acute in the  design of clockless delay-insensitive (DI) circuits, 
leads {one to reconsider the classical descriptions of time and of} the resulting 
order and  causal relations satisfied by logical operations. 
This {questioning} meets the criticisms of classical spacetime formulated by Einstein when founding relativity theory and is answered by relativistic conceptions of time and causality. 
{Applying this approach to clockless circuits and considering the trace formalism, we rewrite Udding's rules which characterize communications between DI components. 
We exhibit their intrinsic relation with relativistic causality.
For that purpose, we introduce relativistic generalizations of traces, called R-traces, which
provide a pertinent description of communications and compositions of DI components.}

\end{abstract}

\section{Introduction}

Regular and important advances in semiconductor technology allow more and more complex processors to be designed.
Meanwhile, progress in computing capacity is accompanied by a need for good {representations}
of physical devices, which are required to remain pertinent at smaller and  smaller scales.
 This trend naturally increases the number of physical properties to be accounted for when implementing logical devices. 
Appropriate {models}  become  mandatory prerequisites for developing
new techniques.
This particularly is the case when adopting the approach of distributed systems or clockless circuits.  

Obviously, time plays an important part in the operation of computing systems.
 The significant effects induced by time delays in modern
electronic circuits point at limits that one rapidly reaches in a classical framework when looking for
physically precise models. 
The only way to go beyond these limits  is to adopt a new framework, able
to match physical reality at a more fundamental level.  

Practical difficulties met in disseminating and managing clock signals over complex designs 
have made the approach of clockless circuits particularly interesting and promising.
But removing the clock does not mean that designs are freed from all the constraints associated with time.
Time is by itself a very complex notion, which is hardly captured in totality, even in most   
advanced physical theories. Designing circuits around a single clock {appears as a convenient way}
 to implement known and classical properties related to time. 
In {the} absence of a clock, other solutions have to be found 
to ensure the proper time evolution  of a circuit.
 To this end, one must reconsider 
the classical models of time that have been useful when designing existing electronic circuits. 

{Important} progress in the control of time constraints in clockless circuits 
has been made with the notion of {\it delay insensitivity}
\cite{FRW,udding-these,snepscheut85,Ebergen-thesis,verhoeff94}.
By removing any dependence on  time lapses due to signal propagation, one becomes able,
with a simple set of rules, to ensure the correct functioning of clockless logical devices.
With this efficient trick comes a new {concept} of the way distributed systems
can synchronize their actions, by means of communication, in order to realize specified computations.

Remarkably, this approach follows, in the context of logical devices,
 a line which is very close to that followed by Einstein, in the general context of physics,
 when introducing the founding concepts of relativity theory \cite{Einstein-zur,EinsteinColl}.
The necessity of constructing a new framework emerges from the remark that no physical system exists
that can deliver time over all space simultaneously. Physical time, as it can be observed,
is necessarily a spatially localized quantity. To promote time to a physical quantity that
can be shared by remote observers, propagation signals must be used. 
{Resulting} propagation delays 
must be taken into account when synchronizing different ``local times"{, that} {profoundly}
change the relations between space and time assumed in classical theories.   

A change of the notion of time has major consequences for the expression of causality 
\cite{russell-ABC,russell-Brit}. As causality clearly depends on a notion of order
in time, a drastic change in the possibility of time ordering, as raised by the relativistic framework, 
implies a revision of the  dependence of all operations, including logical ones,  on causal relations.

The effect of the  change in the properties of time induced by relativity
on distributed systems has already been discussed in the literature
(\cite{LamportACM78,Lamport1986,Mattern,matherat2003}).
{Many formalisms also exist that describe computing concurrency and have provided  efficient
models for designing and analyzing asynchronous circuits. On one hand,
Petri networks \cite{petri62,murata89} and their  interpretation called signal transition graphs (STG) 
\cite{ros-yak:85,chu-leun-wa:85} have provided powerful representations of partial orders and
 causal relations in logic circuits.
 On the other hand,
trace semantics, communicating sequential processes (CSP) \cite{Hoare-CSP}, and related approaches
\cite{udding-these,snepscheut85,Ebergen-thesis,Mazurkiewicz,DBLP:conf/cav/ProbstL90}, are preferred in modelling complex systems.
A formalism sharing both qualities would be particularly helpful for efficiently
simulating and implementing causal relations in asynchronous circuits of high complexity.}

{We shall show here that a closer account of physical causality, based on the relativistic  
notion of time ordering, provides a way to modify
 the formalism of traces so that causal relations can be represented in a natural way. 
This representation should make them easier to implement in arbitrarily complex designs.
In order to fulfill that aim, a key step is to give communications between components a representation that
naturally satisfies the constraints imposed by relativistic time orders.  
Remarkably, Udding's rules \cite{udding-these}, which are used to define communications
within DI systems, appear to suit this purpose.  Rewriting these rules in terms of R-traces, we shall exhibit their intrinsic connection with relativistic causality.}

{In the first two sections we recall} the profound changes brought by the relativistic framework
to the notions of time, time-ordering and causal relations.
In the following sections, we introduce generalizations of traces, called R-traces, and use  them to rewrite 
Udding's rules. {We also briefly describe how R-traces map onto the usual traces
in a classical environment.
In conclusion, we point at some properties and other potential applications of R-traces.}

\section{Relativity and time-ordering}
\label{relativity}

In this section, we  briefly recall the  properties of time according to relativity theory
\cite{Einstein-zur,EinsteinColl,russell-ABC,russell-Brit}, in contrast to those which are implicitly  assumed 
  in a classical framework.
We discuss the new status given by relativity theory to simultaneity in time,
 and  the notion of order in time that follows.
Consequently, 
 the notion of causality used in a relativistic framework significantly differs from its classical analog.

Although time and space can be treated as distinct concepts in classical theory,
this can no longer be the case in a relativistic framework, where  these notions  cannot  be considered 
as being {independent}. 
To be precise, within a classical framework, time is not affected by a change of frame (or observer).
For instance, coordinates $(t,x)$ in an inertial frame are related to coordinates $(t^\prime,x^\prime)$
in another frame, with relative velocity $v$, according to the usual Galilean transformation  
(for simplicity, expressions are written here for a single spatial dimension):
\begin{eqnarray}
x^\prime & =  & x-vt       \nonumber\\
t^\prime & =  & t               \label{class}
\end{eqnarray}
On the other hand, the same change of inertial frame within a relativistic framework
introduces a dependence of time on spatial positions  (light propagates at a finite velocity $c$):
\begin{eqnarray}
x^\prime & = & \frac{x-vt}{\sqrt{1-\frac{v^2}{c^2}}}                       \nonumber\\
t^\prime & = & \frac{t - \frac{v}{c^2} x}{\sqrt{1-\frac{v^2}{c^2}}}         \label{relat}
\end{eqnarray}
This fundamental property is in fact a direct consequence of the definition of time in terms
of physical observables {and of the existence of a maximum
propagation speed, $c$}.
No physical system exists that can provide time simultaneously
 within a whole extended area in space. Time can only be obtained from a clock 
 locally, that is at a given place in space \cite{Einstein-zur,EinsteinColl}.
Time can then only be defined over all space by comparing clocks located at different spatial positions.
These comparisons are made by means of propagating signals.
As a consequence of the delays due to propagation at finite speed (even at light velocity $c$)
the notion of simultaneity is frame dependent.    
 For instance, in contrast to its classical analog
(\ref{class}), the relativistic transformation of time (\ref{relat}) shows that 
two events $A$ and $B$ that occur simultaneously but at different places  in a given frame, 
loose their simultaneity when they are seen in a relatively moving frame.
\begin{eqnarray}
\label{simultaneity}
{\rm{classical}}\quad (1): \qquad &&t_B=t_A, \quad x_B\neq x_A \quad \Rightarrow \quad t_B^\prime=t_A^\prime   \nonumber\\
{\rm{relativistic}}\quad(2):  \qquad &&t_B=t_A, \quad x_B\neq x_A \quad \Rightarrow \quad t_B^\prime\neq t_A^\prime
\end{eqnarray}

The existence of an absolute global time, independent of space, is ruined by the impossibility of  defining simultaneity
 over all space in a frame-independent way. This however excludes neither the possibility of defining
simultaneity, but in a frame-dependent way, nor the existence of frame-independent properties of time. 
The last possibility appears to be easier to realize.
Following that option, one remarks that
expressions (\ref{relat}) describe the transformations of space and time under the action of the Lorentz group.
These transformations generalize to spacetime the action of the group of spatial rotations.
It corresponds to transformations that preserve the spacetime interval $I_{AB}$ between two events
$A$ and $B$.
 \begin{eqnarray}  
\label{interval}
I_{AB} = c^2 (t_A - t_B)^2 -  (x_A - x_B)^2 
\end{eqnarray}
Spacetime intervals (\ref{interval}) are invariant under frame transformations {(\ref{relat})}
{and allow one to define spacetime relations and  subsets that are independent of the observer.
In particular, a null spacetime interval ($I_{AB}=0$) characterizes a pair of events $(A,B)$ that can be joined by
a motion at the maximum speed $c$, {i.e.} which lie on the same light ray. 
The light rays that are incident on a same event $A$ define a subset $\C_A$,
the light cone from event $A$ (see Equation (\ref{interval})).
\begin{eqnarray}
\C_A \equiv \{B: \ I_{AB}=0\}
\end{eqnarray}
$\C_A$ bounds the part of spacetime which can be reached by signals 
sent or received at $A$: it  
realizes a partition of spacetime into three subsets}.
 \begin{eqnarray}
\label{light-cones}
&&\C_{A_>} \equiv \{B: \ I_{AB} > 0\}\nonumber\\
&&\C_A \equiv \{B: \ I_{AB}=0\}\nonumber\\
&&\C_{A_<} \equiv \{B: \ I_{AB}< 0\}
\end{eqnarray}
The interior of a light cone $\C_{A_>}$ consists of all spacetime events $B$ that are 
separated  from  event $A$ by a {time-like} interval ($I_{AB}>0$), 
while the exterior of the same light cone $\C_{A_<}$ consists of 
all events $B$ that are separated from event $A$ by a {space-like} interval ($I_{AB}<0$).
The interior of light cone $\C_{A_>}$ furthermore consists of two disjoint parts which describe
the \textit{future} of event $A$ ($\C_{A_>}^+$), and the \textit{past} of event $A$  ($\C_{A_>}^-$).
\begin{eqnarray}
\label{future_past}
&&\C_{A_>}^+ \equiv \{B: \ I_{AB} > 0, \ t_B-t_A>0\}\nonumber\\
&&\C_{A_>}^- \equiv \{B: \ I_{AB} > 0, \ t_B-t_A<0\}
\end{eqnarray}
All the latter properties are frame-independent. 
Let us note that as opposed to the interior $\C_{A_>}$, 
the exterior of a light cone $\C_{A_<}$ cannot be partitioned in a frame-independent way
 into two parts corresponding to a positive ($t_B-t_A>0$) or negative ($ t_B-t_A<0$) time interval. 
Indeed, for events $B$ in $\C_{A_<}$, the sign of the time interval
$t_B-t_A$ depends on the observer.
These properties exhibit a major difference due to the {relativistic}  framework:
given a particular event $A$, there exists a subset of events $B$ that have a definite position in time
with respect to $A$ ($\C_{A_>}$), but also a large subset of events $B$ that cannot have 
a definite position in time relative to event $A$ ($\C_{A_<}$).	 

Equations (\ref{future_past}) show that, although two arbitrary events cannot be ordered in time, due 
to the frame dependence of the simultaneity property (\ref{simultaneity}), nonetheless 
all events belonging to $\C_{A_>}^+$ are posterior to $A$ while all events belonging to
$\C_{A_>}^-$ are anterior to $A$, both properties being frame-independent.
One remarks that the existence of a time order between two events is a symmetric property:
\begin{eqnarray}
\label{causal_relation}
&&B \in \C_{A_>}^\pm \quad \Leftrightarrow \quad A \in \C_{B_>}^\mp\\
&&C \in \C_{A_<} \quad \Leftrightarrow \quad A \in \C_{C_<}\label{acausal_relation}
\end{eqnarray}
Two events $A$ and $B$ satisfying conditions (\ref{causal_relation}) are ordered in time 
and can thus be causally related. Moreover, this property is easily seen to be transitive.
\begin{eqnarray}
\label{transitivity}
B \in \C_{A_>}^\pm \ \wedge \ C \in \C_{B_>}^\pm \quad \Rightarrow \quad C \in \C_{A_>}^\pm
\end{eqnarray}
In other words, although not allowing a total time ordering of events, the relativistic framework
still allows one to define a partial time-ordering that does not depend on the observer.
This property is necessary and sufficient for formulating causality in physics.

 \begin{figure}[tp] 
   \centering

\begin{picture}(0,0)%
\includegraphics{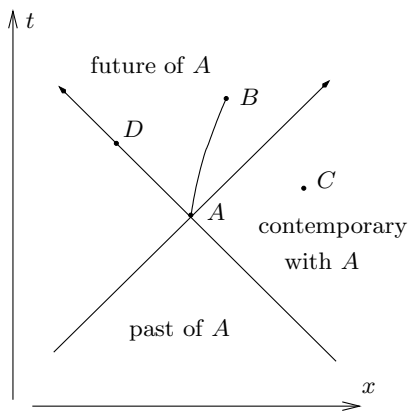}%
\end{picture}%
\setlength{\unitlength}{3522sp}%
\begingroup\makeatletter\ifx\SetFigFont\undefined%
\gdef\SetFigFont#1#2#3#4#5{%
  \reset@font\fontsize{#1}{#2pt}%
  \fontfamily{#3}\fontseries{#4}\fontshape{#5}%
  \selectfont}%
\fi\endgroup%
\begin{picture}(2874,2844)(1129,-2758)
\put(1936,-826){\makebox(0,0)[lb]{\smash{{\SetFigFont{9}{10.8}{\rmdefault}{\mddefault}{\updefault}{\color[rgb]{0,0,0}$D$}%
}}}}
\put(2746,-601){\makebox(0,0)[lb]{\smash{{\SetFigFont{9}{10.8}{\rmdefault}{\mddefault}{\updefault}{\color[rgb]{0,0,0}$B$}%
}}}}
\put(3286,-1186){\makebox(0,0)[lb]{\smash{{\SetFigFont{9}{10.8}{\rmdefault}{\mddefault}{\updefault}{\color[rgb]{0,0,0}$C$}%
}}}}
\put(2521,-1411){\makebox(0,0)[lb]{\smash{{\SetFigFont{9}{10.8}{\rmdefault}{\mddefault}{\updefault}{\color[rgb]{0,0,0}$A$}%
}}}}
\put(1981,-2221){\makebox(0,0)[lb]{\smash{{\SetFigFont{9}{10.8}{\rmdefault}{\mddefault}{\updefault}{\color[rgb]{0,0,0}past of $A$}%
}}}}
\put(1711,-376){\makebox(0,0)[lb]{\smash{{\SetFigFont{9}{10.8}{\rmdefault}{\mddefault}{\updefault}{\color[rgb]{0,0,0}future of $A$}%
}}}}
\put(3061,-1756){\makebox(0,0)[lb]{\smash{{\SetFigFont{9}{10.8}{\rmdefault}{\mddefault}{\updefault}{\color[rgb]{0,0,0}with $A$}%
}}}}
\put(2881,-1501){\makebox(0,0)[lb]{\smash{{\SetFigFont{9}{10.8}{\rmdefault}{\mddefault}{\updefault}{\color[rgb]{0,0,0}contemporary}%
}}}}
\put(1261,-61){\makebox(0,0)[lb]{\smash{{\SetFigFont{9}{10.8}{\rmdefault}{\mddefault}{\updefault}{\color[rgb]{0,0,0}$t$}%
}}}}
\put(3601,-2626){\makebox(0,0)[lb]{\smash{{\SetFigFont{9}{10.8}{\rmdefault}{\mddefault}{\updefault}{\color[rgb]{0,0,0}$x$}%
}}}}
\end{picture}%
   \caption{A spacetime diagram}
   \label{esp-tmp}
\end{figure}
 
Time-order relations are advantageously {illustrated} on spacetime diagrams. In Figure \ref{esp-tmp},
$D$ represents an event with vanishing interval with respect to event $A$,
 {i.e.} located on the light cone $\C_A$  from $A$.
Note that, up to now, we have used the term `event' in accordance with its physical meaning,
{i.e.} to denote a point in spacetime. 
In contrast, in the context of logic devices,
the term `event' usually denotes a transition or a pulse,
 propagating through a wire or an electromagnetic
channel {\cite{SUTH89}}. According to previous discussions, 
the latter rather corresponds to a ray, connecting points located on a same light cone
 (Figure \ref{esp-tmp}). But, it is easily seen that, when {sent or received}, 
such a logical  `event' also corresponds to a transition that is localized in time,
 and which can thus be identified
with a spacetime `event' \cite{russell-ABC,russell-Brit,russellAM,SUTH89}. 
These two characteristic properties, namely localization in time and 
spreading over space, are precisely the ones that are necessary for synchronizing actions 
performed at different locations in space. 
{They also allow the local notion of `logical event', 
defined as a logical action occurring in a  device   
in coincidence with a spacetime `event', to be extended to communications, {i.e.}:
logical actions occurring in different devices in a correlated way. 
In the latter case, `logical events' are associated with propagations rather than points in spacetime.
As previously discussed, they must respect the constraints fixed by the existence of a bound on propagation speeds,
hence the light cone partitions of spacetime.
They may alternatively be seen as a minimal description, regarding logical properties, of events occurring at
the boundary between two devices with different spatial locations.} 
For simplicity and when no confusion may arise, we shall
in the following use the same term `event' for denoting both concepts.

Figure \ref{esp-tmp} also shows an event $B$ that lies in the future light cone of $A$ ($B \ \in \ \C_{A>}^+$). This property is equivalent to the possibility for 
a physical system to travel from event $A$ to event $B$ with a speed that is lower than light velocity $c$,
the largest allowed velocity. 
Event $B$ can then be said to {occur} \textit{after} event $A$.
In contrast, as no definite time order can be attributed to an event  $C$ lying outside of the light
 cone from $A$ ($C \ \in \ \C_{A<}$, $I_{AC} < 0$ is {space-like}),
$C$ is said to be `contemporary with $A$'. 

These properties can be used to express relativistic causality.
Two actions {occurring} at different locations  in spacetime can be causally related
 only if the two events $A$ and $B$ at which they occur are ordered in time, {i.e.} satisfy relation
(\ref{causal_relation}). In that case, these two actions may produce a result that depends on the
time-order of events $A$ and $B$. Conversely, two actions {occurring} at two events $A$ and $C$
that are contemporary, {i.e.} $A$ and $C$ satisfy (\ref{acausal_relation}),
produce a result that cannot depend on their order, as the latter cannot be identified with 
an order in time. 
One deduces that the result of two actions {occurring} at two different events $A$ and $B$
can depend on the order of these actions if and only if $A$ and $B$ can be ordered in time
(\ref{causal_relation}).
Otherwise, two actions {occurring} at events satisfying Equations (\ref{acausal_relation}) 
 must be considered to be independent of each other and hence cannot satisfy a causal relation.

\section{Causality and time delays}
 
Computation reduces, in present realizations, to  chaining elementary
operations performed on {Boolean} variables,
 linking these operations through a finite set of prescribed logical relations.
This occurs in logic devices performing successive operations
ruled by a clock or in asynchronous (clockless) circuits as well. In all cases, 
specifications are satisfied by linking the prescribed logical rules  
to causal relations constraining the physical systems in which they are realized.       
 Not surprisingly, logic devices can be  designed in a satisfactory way only when 
 an appropriate representation
of causal relations is available. 

Modern  logic devices are implemented in highly integrated components,
pushing the limits of transport properties. 
Due to clocks operating at high frequencies and despite small dimensions, the 
limits affecting the propagation of signals within a chip, due to {bounded propagation speeds},
cannot be ignored. A major consequence, as discussed in previous section, is a breakdown of the classical (Newtonian or Galilean)  representation of time-ordering, pointing at the necessity of a 
genuinely relativistic representation. 
Obviously, the same remarks apply to distributed systems--to systems made of components with large separations (measured as propagation times)  when compared with their internal period \cite{LamportACM78,Lamport1986,Mattern}.
Distributed systems or asynchronous circuits can be seen to involve  
two different kinds of logical processes:
\begin{itemize}
\item transitions associated with  changes of state of a local component,
\item {communications} associated with exchanges of data between components.
\end{itemize}  
Both kinds of processes must be considered when defining the specifications of  distributed systems 
or asynchronous circuits and when setting their communication protocols. 
Moreover, these logical processes should be associated with implementations corresponding
to two different kinds of events,
this notion being understood according to the two different conceptions discussed in the previous section.

The implementation of logical processes relies on the existence of causal relations, hence on
the possibility of ordering in time, the corresponding events.
More precisely, the time order of two events  becomes an objective property
 as soon as each event belongs to the interior of the
other's light cone (\ref{causal_relation}), or else as soon as they define a {time-like} interval
(\ref{light-cones}).
Two different kinds of  {time-like} intervals can be seen to be
involved in the definition of distributed systems or asynchronous circuits:
\begin{itemize}
\item intervals between two events characterizing the successive states of one local component;
\item intervals between two events characterizing the emission and reception of communication signals.
\end{itemize}
The fact that the second case also corresponds to a {time-like} interval rather than to a {light-like} interval
is due to the fact that, depending on the technology and implementation choices, 
signals may propagate  at a maximum {speed} {smaller than or equal to, the light velocity}.
This entails no fundamental consequences, as all
 properties due to the existence of a maximum {propagation speed} still hold.  

Figure \ref{esp-tmp} shows two events $A$ and $B$, with $B$ {occurring} after $A$ ($B \in \C_{A>}^+$).
Although different observers may attribute different dates to these two events,
 corresponding to different time delays undergone by a signal propagating from $A$ to $B$,
 all observers will nonetheless agree on the {time-like} character of the interval $I_{AB}$ ($I_{AB}>0$)
and on the time-order of the two events ($B\in \C_{A>}^+, \ A\in \C_{B>}^-$).
{The corresponding physical time delay coincides with the {\it delay} 
 accounted for  when designing electronic circuits.
 As opposed to the time delay itself, 
the {time-like} character of the interval between two events $A$ and $B$ and their time order (here $B$ {\it after} $A$), are
objective properties.
As illustrated by this simple example, the time-ordering properties of logical events, when physically implemented
in logic devices, correspond to {\it delay insensitive} (DI) properties.}

{When implementing logical operations in physical devices, 
the dependence of physical causality on time-ordering imposes constraints that tend to limit their performance.
In practice, one must either control time-orders, hence time delays, at the level required
by the operation frequency and size of the device, or design processors so that their operation
can resist arbitrary variations of time delays. These options are not exclusive and solutions adopted in
existing processors take advantage of both possibilities. Timing methods significantly help improve the
performance of clocked circuits, and nowadays, processors routinely include self-timed asynchronous parts to 
benefit from a locally enhanced level of performance \cite{stevens:tvlsi03}. 
The increasing cost in both complexity and power consumption, entailed by monitoring time lapses over
highly integrated circuits, however leads one to favor approaches that can escape, as much as possible, time delay constraints. The formalization of delay insensitivity (DI)
\cite{FRW,udding-these,Ebergen-thesis,verhoeff94} provides a powerful conceptual
framework for designing asynchronous circuits. Confronting this approach
to the relativistic background of physical processes,
we shall try in the following to understand DI rules as the result of general and fundamental
constraints imposed by physical causality on the implementation of logic devices. 
}

\section{Partial order relations} \label{causal}

We now focus  on logical operations and on their implementations as processes 
 in physical devices. In general, the result of two successive logical operations
depends on the order in which they are performed. At the level of physical processes,
the ordering of logical operations is given by the time-ordering of  the events at  which they occur.
 For that reason, we shall identify
 in the following, the  time-ordering of logical operations  with the ordering of the corresponding events.
 Upper case letters denoting events will just be replaced by lower case letters
when denoting logical operations.  
We also introduce some simplifying notations to represent ordering when expressed on logical operations
rather than events.

\textbf{Definition}: When the event associated with a logical operation $b$ is in the future of 
the event associated with a logical operation $a$, we say that $a$ and $b$ are ordered (in time) 
and write this relation with the following three symbols
\begin{eqnarray}
\label{time_order}
{a \vartriangleleft b}
\end{eqnarray}
This relation is \textit{irreflexive}, \textit{anti-symmetric} and \textit{transitive}. We say that $a$ is \textit{before} $b$ or $b$ is \textit{after} $a$.

\textbf{Definition}: When two logical operations  $a$ and $c$ are such that neither relation $a \vartriangleleft c$ 
nor relation $c \vartriangleleft a$ holds, we say that $a$ and $c$ are  \textit{contemporary} and  write
$$a \bowtie c$$
This relation is \textit{symmetric} but not  \textit{transitive}, for  $a \bowtie c$ and $c \bowtie b$ can both hold  without  $a \bowtie b$ being satisfied (see Figure \ref{esp-tmp}). 

Relations $ \vartriangleleft $ and $ \bowtie $, which are  based on relations between events, admit 
simple repesentations on spacetime diagrams. 
Let us note that  relations $ \vartriangleleft, \bowtie$ can also be seen as a particular case of the relations $\longrightarrow, \dashrightarrow$  introduced by Lamport \cite{Lamport1986}
to provide a general formalization of  logical time in distributed computations. 
These definitions however differ in an important way: $ \vartriangleleft $ and $ \bowtie $ refer here to
an ordering with respect to `physical time', although $\longrightarrow, \dashrightarrow$ refer to a constructed `logical time' \cite{Lamport1986}. 
One could alternatively say that the choice to {model} implementations of logic devices
at the level of  electronic components
leaves less freedom for a {representation} of time.

As already stated, the expression of causality in physics strongly depends on the notion of time ordering.
It should be clear however that causal relations do not reduce to time-order relations.
Obviously, two physical actions, in particular two logical operations, can happen to be ordered in time without being
causally related. In other words, two logical operations can only be causally related if they are 
ordered in time, the latter condition being necessary but not sufficient. 
For that reason, {a further notation} must be introduced to describe causal relations.

\begin{figure}[tp] 
   \centering
\begin{picture}(0,0)%
\includegraphics{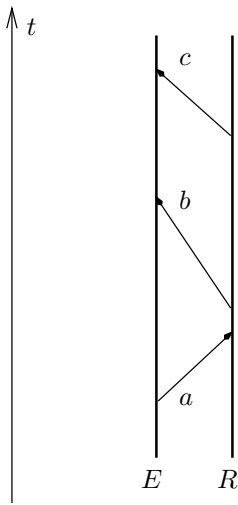}%
\end{picture}%
\setlength{\unitlength}{4144sp}%
\begingroup\makeatletter\ifx\SetFigFont\undefined%
\gdef\SetFigFont#1#2#3#4#5{%
  \reset@font\fontsize{#1}{#2pt}%
  \fontfamily{#3}\fontseries{#4}\fontshape{#5}%
  \selectfont}%
\fi\endgroup%
\begin{picture}(1369,2994)(859,-3448)
\put(1666,-3346){\makebox(0,0)[lb]{\smash{{\SetFigFont{10}{12.0}{\rmdefault}{\mddefault}{\updefault}{\color[rgb]{0,0,0}$E$}%
}}}}
\put(2116,-3346){\makebox(0,0)[lb]{\smash{{\SetFigFont{10}{12.0}{\rmdefault}{\mddefault}{\updefault}{\color[rgb]{0,0,0}$R$}%
}}}}
\put(991,-646){\makebox(0,0)[lb]{\smash{{\SetFigFont{10}{12.0}{\rmdefault}{\mddefault}{\updefault}{\color[rgb]{0,0,0}$t$}%
}}}}
\put(1891,-2851){\makebox(0,0)[lb]{\smash{{\SetFigFont{10}{12.0}{\rmdefault}{\mddefault}{\updefault}{\color[rgb]{0,0,0}$a$}%
}}}}
\put(1891,-1681){\makebox(0,0)[lb]{\smash{{\SetFigFont{10}{12.0}{\rmdefault}{\mddefault}{\updefault}{\color[rgb]{0,0,0}$b$}%
}}}}
\put(1891,-826){\makebox(0,0)[lb]{\smash{{\SetFigFont{10}{12.0}{\rmdefault}{\mddefault}{\updefault}{\color[rgb]{0,0,0}$c$}%
}}}}
\end{picture}%

   \caption{Partial causal order}
   \label{ordre-causal}
\end{figure}

\textbf{Definition}: To express ``$a$ is the cause of $b$", we write
\begin{eqnarray}
\label{causal_order}
a \rightarrow b
\end{eqnarray}
{The constraint between causal and time-order relations may then be written
$$(a \rightarrow b) \quad \Rightarrow \quad (a \vartriangleleft b)$$
The converse implication does not hold and a} configuration showing the difference between causal and time-order relations is illustrated 
in Figure \ref{ordre-causal},
where $E$ and $R$ represent two components, and $a$, $b$ and $c$ communications  between $E$ and $R$.
 Figure \ref{ordre-causal} illustrates a case where $a$ is the cause of $b$ and $c$, 
with no relation of causality between $b$ and $c$.
This gives an example of relation $b \vartriangleleft c$ holding  without relation $b \rightarrow c$ being true.

The causal situation illustrated in Figure \ref{ordre-causal} is also compatible with another  configuration in time, $a \vartriangleleft  (b \bowtie  c)$. 
Such a configuration  happens when the environment $E$ is placed farther on the  left.
This results from the fact that  when  $c$ is emitted by $R$ one does not know, in the absence of acknowledge, 
whether  $b$ has already been received by $E$. The time-order realized between operations $b$ and $c$ 
in Figure \ref{ordre-causal} is an example of a {\it delay sensitive} relation.
{ The causal order specified in Figure \ref{ordre-causal} can be realized with different time orders},
$(a \vartriangleleft b \vartriangleleft c)$, $(a \vartriangleleft c \vartriangleleft b)$ or $(a \vartriangleleft (b \bowtie c))$.
This example shows that a total time-order is not necessary to satisfy all causal relations required by
specifications. On the other hand,  {delay insensitivity forbids}   situations 
such as the one illustrated in Figure  \ref{ordre-causal}. Delay insensitivity expresses
 the compatibility of arbitrary timing configurations with the causal order specified by  a computation.
Hence,  it appears as a general and simple way to account for the arbitrariness in time ordering that remains once
causal orders are satisfied.
Using only DI circuits then allows one to replace the causal order $\rightarrow$
defined in Equations (\ref{causal_order}) by the 
time-order $\vartriangleleft$ defined in Equations (\ref{time_order}) when describing the logical constraints
associated with specifications.

\section{Structures of relativistic traces}

In this section, we describe a formalism of relativistic traces (R-traces) for electronic circuits, 
generalizing standard traces {used for DI systems}
\cite{udding-these,snepscheut85,Ebergen-thesis,verhoeff94}
so as to account
for the orders defined in the previous section. 

For simplicity, in the following, we identify logical operations with the events at which they occur.
Symbols consist of lower case letters of the beginning of the alphabet ($a$, $b$, $c$, ...), thus denoting events, while referring to logical operations. 
The concatenation operator ``$;$" (concatenation of symbols or of strings of symbols) is replaced by operators describing time-ordering, which we choose to be  ``$\vartriangleleft $" and ``$\bowtie $". We also add parentheses  ``$($" and ``$)$".
{Note} that R-traces do not represent the causal order (denoted by  ``$\rightarrow$") which  will appear later.

R-trace {\it structures} for a component are defined as triples 
$S = \; <\mathbf{i}S, \mathbf{o}S, \mathbf{t}S>$, where:
\begin{itemize}
\item $\mathbf{i}S$ is a finite set of symbols denoting propagation events from the environment to the component (\textit{inputs}),
\item $\mathbf{o}S$  is a finite set of symbols denoting propagation events from the component to the environment (\textit{outputs}),
\item $\mathbf{t}S$ denotes a set (finite or not) of finite length strings (named \textit{R-traces}) written with symbols in the set
{$\mathbf{a}S$, where $\mathbf{a}S$ (\textit{alphabet} of $S$) is\\
$\mathbf{i}S \cup \mathbf{o}S \cup \{ \vartriangleleft, \bowtie, (, ) \}$} and obeys the following syntax.
\end{itemize}

R-trace structures are denoted by upper case letters  ($R$, $S$, $T$, ...) and \textit{R-traces} are denoted by the lower case letters ($s$, $t$, $u$, ...) of the end of the alphabet.
Moreover, the following properties are assumed to hold.
\begin{itemize}
\item The empty R-trace, denoted  by ``$\varepsilon$", belongs to the structure $S$: $ \varepsilon \in \mathbf{t}S$.
\item  A single symbol can be an R-trace, for example: $ a \in \mathbf{t}S$.
\item An R-trace $s$ can be extended with a new symbol $a$  when $s$ is entirely in the past  of $a$ ($s \vartriangleleft a \in \mathbf{t}S$).
\item  By definition: $\vartriangleleft\ \vartriangleleft\  = \  \vartriangleleft  $ and for any  R-trace $s$, $s\vartriangleleft\  = s = \ \vartriangleleft s $.
\item An event symbol in a R-trace can be replaced by two event symbols with no time order between them; this will be written: $(a \bowtie b)$.
\item By definition: $a = (a)= (\vartriangleleft a)= (a\vartriangleleft) = (a \bowtie)= (\bowtie a)$.
\item When an order symbol ``$\vartriangleleft $" precedes parentheses, each symbol within the parentheses denotes an event that is in the  future of the sequence preceding the order symbol. For example, $s \vartriangleleft (a \bowtie b)$ means that, both and independently, $s$ is before $a$ and $s$ is before $b$.
\item When an order symbol ``$\vartriangleleft $" follows parentheses, each symbol within parentheses  denotes an event that is in the past of the sequence following the order symbol.
 For example, $(a \bowtie b) \vartriangleleft t $ means that $t$ is both after $a$ and after $b$.
\item In an R-trace, an event symbol can be replaced by an R-trace, put inside parentheses. For example, 
 $s \vartriangleleft ((a \vartriangleleft b) \bowtie c) \vartriangleleft t $ is an R-trace. This R-trace means that $s$ is before $a$ and before $c$, and that $t$ is after $b$ and after $c$.
\item Associativity follows:
$(a \bowtie (b \bowtie c)) \equiv ((a \bowtie b) \bowtie c)$, even if the relation ``$\bowtie $" is non-transitive, as this expression means that  $a$, $b$ and  $c$ can only satisfy a time-order relation with symbols  lying outside the { parentheses.}
\end{itemize}

{As} an example, the following R-trace can be  part of the history of a Muller C-element
\begin{equation}
(a \bowtie b) \vartriangleleft c \vartriangleleft a \vartriangleleft b \vartriangleleft c \vartriangleleft b \vartriangleleft  a \vartriangleleft  c \vartriangleleft (a \bowtie b) \vartriangleleft c           \label{c-muller}
\end{equation}

\subsection{Operations on R-Traces}

We define the \textit{concatenation} of two R-traces $t$ and $u$ as $t \vartriangleleft u$.
This operation will be extended to concatenation of sets of R-traces. For example, $\mathbf{t}R \vartriangleleft \mathbf{t}S$ denotes the set of R-traces formed by concatenating one R-trace from $\mathbf{t}R$ and one R-trace from $\mathbf{t}S$.

We define the \textit{star} operation of R-trace $t$, written $ [t]^*$, as the set of
 finite numbers of concatenations of $t$
$$ [t]^* = \{ \varepsilon, t, t \vartriangleleft t, t \vartriangleleft t \vartriangleleft t, t \vartriangleleft t \vartriangleleft t \vartriangleleft t, ...\}$$
Applied to a set of R-traces, the \textit{star} operator produces the set of all traces obtained 
from  a finite number of concatenations of R-traces of the set.

We define the \textit{prefix} operator in the following way.
Let $t$ be an R-trace such that {$t \in \mathbf{t}S$}; the string $u$ is a prefix of $t$ if {$u \in \mathbf{t}S$} and if there exists {$v \in \mathbf{t}S$} such that $t = u \vartriangleleft v$. We will also consider that $\varepsilon$ and $t$ are prefixes of $t$.
The $\mathbf{pref}$ operator applied to $t$ produces the set of all prefixes of $t$. Applied to a set of traces, it produces the set of prefixes of all R-traces of the set.

Let  $t$ be an R-trace, and $A$ a set of symbols including
 $\{ \vartriangleleft, \bowtie, (, ) \}$. 
We define $t \downarrow A$, called the
\textit{projection} of $t$ on $A$, as the  R-trace obtained by only keeping symbols of $t$ belonging to $A$.
\begin{itemize}
\item If $t = a$ and $a \in A$, then $t \downarrow A = a$
\item If $t = a$ and $a \notin A$, then $t \downarrow A = \varepsilon$
\item If $t = u  v$ and $u \downarrow A  = u^\prime \neq \varepsilon$ and $v \downarrow A = v^\prime \neq \varepsilon$, \\ then $t \downarrow A = u^\prime v^\prime$
\item If $t = u  v$ and $u \downarrow A  = u^\prime\neq \varepsilon$ and $v \downarrow A = \varepsilon$, \\ then $t \downarrow A =  u^\prime$
\item If $t = u v$ and $u \downarrow A = \varepsilon$ and $v \downarrow A = v^\prime\neq \varepsilon$, \\ then $t \downarrow A = v^\prime$
\item If $t = u  v$ and $u \downarrow A = \varepsilon$ and $v \downarrow A = \varepsilon$, \\ then $t \downarrow A = \varepsilon$
\end{itemize}
As an example of such a projection,  R-trace (\ref{c-muller}), projected on $\{ a, c, \vartriangleleft, \bowtie, (, ) \}$  produces the following R-trace
$$a \vartriangleleft c \vartriangleleft a \vartriangleleft c \vartriangleleft a \vartriangleleft  c \vartriangleleft a \vartriangleleft  c$$
which is a concatenation of $ a \vartriangleleft c$ and can be rewritten
$$[a \vartriangleleft c]^4$$
Similar operations  can be defined for the following R-trace structures.
\begin{itemize}
\item Concatenation:
$$R ; S = <\mathbf{i}R \cup \mathbf{i}S, \mathbf{o}R \cup \mathbf{o}S, \mathbf{t}R \vartriangleleft \mathbf{t}S>$$
\item Union:
$$R | S = <\mathbf{i}R \cup \mathbf{i}S, \mathbf{o}R \cup \mathbf{o}S, \mathbf{t}R \cup \mathbf{t}S>$$
\item Star:
$$* [R] = <\mathbf{i}R, \mathbf{o}R, [\mathbf{t}R]^* >$$
\item Prefix:
$$\mathbf{pref}R = <\mathbf{i}R, \mathbf{o}R, \mathbf{pref}[\mathbf{t}R]>$$
(If $\mathbf{pref}R = R$,  $R$ is called ``prefix-closed".)
\item Projection (on the set $A$):
$$R \downarrow A = <\mathbf{i}R \cap A, \mathbf{o}R \cap A, \{t \downarrow A \; | \; t \in \mathbf{t}R \}>$$
\item Weave:
$$
R \| S =  <\mathbf{i}R \cup \mathbf{i}S, \mathbf{o}R \cup \mathbf{o}S, \{t {\in} ({\mathbf{t}R} \cup {\mathbf{t}S}) \; | \; 
t \downarrow \mathbf{a}R \in \mathbf{t}R \ \land \ t \downarrow \mathbf{a}S \in \mathbf{t}S \}>
$$
For instance:   $R = \; < \{a \}, \{c \}, \{ a \vartriangleleft c \} >$ and 
$S = \; < \{b \}, \{c \}, \{b \vartriangleleft c \} >$ 
$$
R \| S =  < \{a, b \}, \{c \},
\{ (a \bowtie b) \vartriangleleft c , a \vartriangleleft b \vartriangleleft c , b \vartriangleleft a \vartriangleleft c \}  >$$
\end{itemize}

\subsection{Component and Environment}

The dialog between a \textit{component} and its \textit{environment} is specified by an R-trace structure $S$ (named \textit{command}), which is non-empty ($\mathbf{t}S \neq \emptyset$), 
and such that $\mathbf{i}S \cap \mathbf{o}S = \emptyset$.
Symbols in $\mathbf{i}S$ denote signals emitted by the environment and received by the component. Symbols in $\mathbf{o}S$ denote signals emitted by the component and received by the environment.

One does not make any distinction between the specification of the component on one hand and 
the specification of the environment on the other hand.
One only specifies the \textit{pair} component-environment, defined by a dialog
 described by a single R-trace structure.

For instance, the dialog between a Muller C-element and its environment is {described} in the following way
\begin{equation}
\mathbf{pref} * < \{a, b \}, \{c \}, 
\{ (a \bowtie b) \vartriangleleft c , a \vartriangleleft b \vartriangleleft c , b \vartriangleleft a \vartriangleleft c \} \label{cmullspec} >
\end{equation}
One can define \textit{atomic commands}  as particular R-trace structures
$$
\begin{array}{ccc}
a? & = &< \{a \}, \emptyset , \{a \} > \\ 
b? & = &< \{b \}, \emptyset , \{b \} > \\
c! & = &< \emptyset , \{c \}, \{c \} >
\end{array}
$$
so that Equation (\ref{cmullspec}) can then be rewritten
$$ \mathbf{pref} * [ (a? \| b?) ; c! ] $$

As a consequence of the definition of the \textit{projection} of R-traces, the logical operation {$ a \| b$ 
corresponds to the three
possible time relations $a \bowtie b$, $a \vartriangleleft b$, and $b \vartriangleleft a$.
The previous notation  ``$\|$"  for a DI-component is compatible with the classical weave notation} (as for example in {\cite{Hoare-CSP,Ebergen-thesis}}), but one must note that it now has a different meaning: R-trace structures are a description of a partial order, the {time-order relation} for events that are propagations
 in relativistic spacetime. 
Let us  remark that propagation delays are already included in this description, {in contrast} to the classical description.
Also,  $a?$ is not an input port of a component, but the propagation of a signal that travels from the environment to the component {(see the discussion at the end of Section \ref{relativity}).}

\section{Relativistic rules}

Udding's rules \cite{udding86} are a formalization of the intuitive idea underlying  DI-systems, which was previously introduced under the form of the Foam Rubber Wrapper (FRW) metaphor  \cite{FRW}. The FRW metaphor discusses how a change of propagation delays can change the order of signals in time, and suggests ways to protect component specifications 
 against such changes in time-ordering.
 Udding translated the notion of FRW into a set of logical constraints, to be imposed {on}
trace structures 
 specifying components, so that specifications become invariant under changes of time-ordering due to propagation delays.

{In the classical framework of traces} \cite{snepscheut85},  a trace is understood to describe 
a total time-order. When the latter changes as a result  of propagation delays, one {interprets} this property by
saying that ``the order at the environment boundary is not the same as the order at the component boundary".
{One can alternatively say in a relativistic framework that
the notion of time-order becomes ambiguous for events at the boundary.} 
This is due to the spatial extension of the boundary between a component and its environment, which
 entails the impossibility of ascribing a definite time-order to events {occurring} in this area.
{In this context, R-traces, which preserve the notion of trace while describing a partial order,
appear as a natural generalization. Meanwhile, R-traces bring another important 
semantic change, as they are not only associated with {events occuring at a single location in space}, but also with propagation events.} 

Adapting Udding's rules, we consider R-traces as constraints imposed on the syntax of component specifications.
R-traces now constrain the specification of a single object {that} is a \textit{component-environment pair} 
{and} express the dialog between a component and its environment, and more precisely  causality relations 
satisfied by  propagations between them. 
Let us note that as they are compatible with time-ordering in relativistic spacetime, these causality relations
 make DI-systems compatible, not only with a change in place\&route on a chip at the design stage, but also with {changes due to process variations or to} motions of components during computation.

In the following, we  rewrite Udding's rules and discuss them  using spacetime diagrams. 
Components and environments (respectively denoted by $R$ and $E$)
are  represented by  vertical lines.
 Propagation events are {illustrated by slanted arrows which represent bounded and non-vanishing speeds}
  (as in Figure \ref{ordre-causal}).
$R$ also denotes the R-trace structure describing the dialog specification of a  pair ($R,E$).
{The} $R_0$ rule is a special case of {the} $R_1$ rule and is  discussed after the latter 
(we keep for rules, the names introduced by Udding \cite{udding86}).

\subsection{{The} $R_1$ Rule}

{The} $R_1$ rule concerns two successive symbols of the same type (two inputs or two outputs). Classically, in a context of total ordering, the two possible relative orders have to be taken into account and have to produce the same result
\begin{equation}
\label{R1_rule}
\begin{array}{c}
\forall s,t \in \mathbf{t}R, \; (a,b\in \mathbf{i}R) \vee (a,b\in \mathbf{o}R)\; ; \\
\quad s a b t \in \mathbf{t}R \; \Leftrightarrow \; s b a t \in \mathbf{t}R
\end{array}
\end{equation}

\begin{figure}[htp] 
   \centering

\begin{picture}(0,0)%
\includegraphics{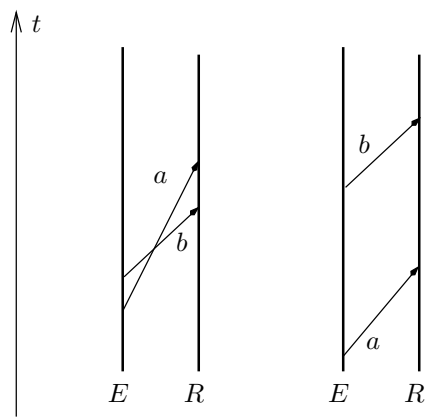}%
\end{picture}%
\setlength{\unitlength}{4144sp}%
\begingroup\makeatletter\ifx\SetFigFont\undefined%
\gdef\SetFigFont#1#2#3#4#5{%
  \reset@font\fontsize{#1}{#2pt}%
  \fontfamily{#3}\fontseries{#4}\fontshape{#5}%
  \selectfont}%
\fi\endgroup%
\begin{picture}(2450,2454)(1129,-2368)
\put(3016,-2266){\makebox(0,0)[lb]{\smash{{\SetFigFont{10}{12.0}{\rmdefault}{\mddefault}{\updefault}{\color[rgb]{0,0,0}$E$}%
}}}}
\put(3466,-2266){\makebox(0,0)[lb]{\smash{{\SetFigFont{10}{12.0}{\rmdefault}{\mddefault}{\updefault}{\color[rgb]{0,0,0}$R$}%
}}}}
\put(1711,-2266){\makebox(0,0)[lb]{\smash{{\SetFigFont{10}{12.0}{\rmdefault}{\mddefault}{\updefault}{\color[rgb]{0,0,0}$E$}%
}}}}
\put(2161,-2266){\makebox(0,0)[lb]{\smash{{\SetFigFont{10}{12.0}{\rmdefault}{\mddefault}{\updefault}{\color[rgb]{0,0,0}$R$}%
}}}}
\put(1261,-61){\makebox(0,0)[lb]{\smash{{\SetFigFont{10}{12.0}{\rmdefault}{\mddefault}{\updefault}{\color[rgb]{0,0,0}$t$}%
}}}}
\put(1981,-961){\makebox(0,0)[lb]{\smash{{\SetFigFont{10}{12.0}{\rmdefault}{\mddefault}{\updefault}{\color[rgb]{0,0,0}$a$}%
}}}}
\put(3196,-781){\makebox(0,0)[lb]{\smash{{\SetFigFont{10}{12.0}{\rmdefault}{\mddefault}{\updefault}{\color[rgb]{0,0,0}$b$}%
}}}}
\put(3241,-1951){\makebox(0,0)[lb]{\smash{{\SetFigFont{10}{12.0}{\rmdefault}{\mddefault}{\updefault}{\color[rgb]{0,0,0}$a$}%
}}}}
\put(2116,-1366){\makebox(0,0)[lb]{\smash{{\SetFigFont{10}{12.0}{\rmdefault}{\mddefault}{\updefault}{\color[rgb]{0,0,0}$b$}%
}}}}
\end{picture}%

   \caption{About {the} $R_1$ rule (two inputs case)}
   \label{R1}
\end{figure}

Let us  discuss this rule using Figure \ref{R1}. On the left, signal $a$ has a lower speed 
than signal $b$  (a lower speed may be the consequence of indirect communications
causing further delays). On the right, signal $b$ is sent at such a date  that it lies in the future of $a$ ($a\vartriangleleft b$). 
In a classical framework, this order appears differently for $E$ and $R$ in the first case (on the left), 
while being identical in the second case (on the right). 
The independence on this relative order of the string following  events $a$ and $b$ justifies
 the classical formulation of {the} $R_1$ rule (\ref{R1_rule}).
On the other hand, within a relativistic framework,
 no time order, hence no causal order, can exist between $a$ and $b$.
Consequently, a relativistic version of {the} $R_1$ rule should read
$$\begin{array}{c}
\forall s,t \in \mathbf{t}R, \; (a,b\in \mathbf{i}R) \vee (a,b\in \mathbf{o}R) \; ; \\
( \quad s \vartriangleleft a \vartriangleleft b \vartriangleleft t \in \mathbf{t}R \quad \vee \quad s \vartriangleleft (a \bowtie b) \vartriangleleft t \in \mathbf{t}R \quad ) \\
\Rightarrow \quad
s \vartriangleleft (a \| b) \vartriangleleft t \in \mathbf{t}R
\end{array}$$
{where $(a \| b)$ is a shorthand notation for all possible time-orders $a
 \vartriangleleft b$, $ (a \bowtie b) $ and $ b \vartriangleleft a$ 
(see the discussion at the end of previous section).}
This is the case that was discussed in Section \ref{causal}. 
When no signal (acknowledge) can inform $E$ that $a$ has been received 
before the emission of  $b$, no consensus can be reached by  $E$ and $R$ on the order of $a$ and $b$.  This is acceptable if the order between $a$ and $b$ has no consequence on  subsequent computations.

\subsection{{The} $R_{0}$ Rule}

{The} $R_0$ rule states that two successive events cannot be identical, {i.e.} transmitted on the same channel and of the same type. This rule aims {to avoid confusing or loosing} events.   		
 Classically, it reads
$$\forall s \in \mathbf{t}R, a\in \mathbf{a}R \; ; \quad s a a \not\in \mathbf{t}R$$
{In relativistic spacetime, one must replace  $saa$ in the $R_1$ rule} by $s \vartriangleleft (a \| a)$.
{How} should one {interpret the rule} $R_0$?
In fact, two events $a$ with no causal relation are independently caused by the {same} preceding event $s$. 
If one insists on repeating an event $a$, the second one has to come after an acknowledgement of the first one, in order to avoid any confusion.
{The}  $R_0$ rule may then be rewritten
$$\forall s \in \mathbf{t}R, a\in \mathbf{a}R \; ; \quad s \vartriangleleft (a \| a) \not\in \mathbf{t}R$$

\subsection{{The} $R_{2}$ Rule}
		
{The} $R_2$ rule concerns two symbols of opposite type (one input and one output). 
In general, these two events may be causally ordered and  this order {may} have a meaning.
No ambiguity arises in that case. This corresponds to $a \rightarrow b$, which we  implement
 by $ a \vartriangleleft b$.
But if these two events are not ordered, or if they are ordered while the two {different} orders
remain possible, then {these} possibilities have to entail the same consequences,  as
propagation delays can change the order. Classically, this rule reads
$$\begin{array}{c}
\forall s, t \in \mathbf{t}R, \; (a\in \mathbf{i}R \wedge b \in \mathbf{o}R)
 \vee (a \in \mathbf{o}R \wedge b\in \mathbf{i}R) \; ; \\
 (s a b \in \mathbf{t}R \wedge s  b a \in \mathbf{t}R) \; \Rightarrow \;  (s a b t \in \mathbf{t}R \Leftrightarrow s  b a t \in \mathbf{t}R)
\end{array}$$

\begin{figure}[htp] 
   \centering

\begin{picture}(0,0)%
\includegraphics{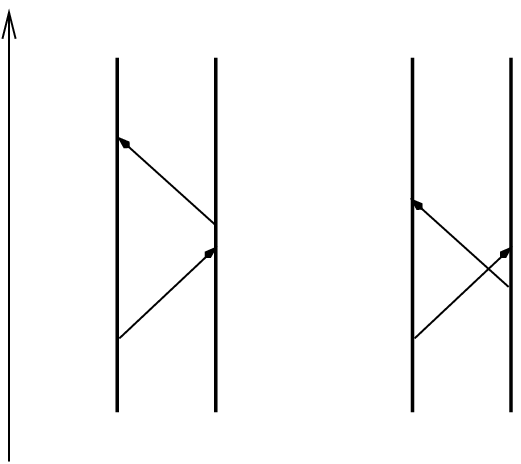}%
\end{picture}%
\setlength{\unitlength}{4144sp}%
\begingroup\makeatletter\ifx\SetFigFont\undefined%
\gdef\SetFigFont#1#2#3#4#5{%
  \reset@font\fontsize{#1}{#2pt}%
  \fontfamily{#3}\fontseries{#4}\fontshape{#5}%
  \selectfont}%
\fi\endgroup%
\begin{picture}(2359,2094)(1129,-2368)
\put(2926,-2311){\makebox(0,0)[lb]{\smash{{\SetFigFont{10}{12.0}{\rmdefault}{\mddefault}{\updefault}{\color[rgb]{0,0,0}$E$}%
}}}}
\put(3376,-2311){\makebox(0,0)[lb]{\smash{{\SetFigFont{10}{12.0}{\rmdefault}{\mddefault}{\updefault}{\color[rgb]{0,0,0}$R$}%
}}}}
\put(3061,-1951){\makebox(0,0)[lb]{\smash{{\SetFigFont{10}{12.0}{\rmdefault}{\mddefault}{\updefault}{\color[rgb]{0,0,0}$a$}%
}}}}
\put(3061,-1141){\makebox(0,0)[lb]{\smash{{\SetFigFont{10}{12.0}{\rmdefault}{\mddefault}{\updefault}{\color[rgb]{0,0,0}$b$}%
}}}}
\put(1576,-2311){\makebox(0,0)[lb]{\smash{{\SetFigFont{10}{12.0}{\rmdefault}{\mddefault}{\updefault}{\color[rgb]{0,0,0}$E$}%
}}}}
\put(2026,-2311){\makebox(0,0)[lb]{\smash{{\SetFigFont{10}{12.0}{\rmdefault}{\mddefault}{\updefault}{\color[rgb]{0,0,0}$R$}%
}}}}
\put(1801,-1861){\makebox(0,0)[lb]{\smash{{\SetFigFont{10}{12.0}{\rmdefault}{\mddefault}{\updefault}{\color[rgb]{0,0,0}$a$}%
}}}}
\put(1261,-466){\makebox(0,0)[lb]{\smash{{\SetFigFont{10}{12.0}{\rmdefault}{\mddefault}{\updefault}{\color[rgb]{0,0,0}$t$}%
}}}}
\put(1801,-961){\makebox(0,0)[lb]{\smash{{\SetFigFont{10}{12.0}{\rmdefault}{\mddefault}{\updefault}{\color[rgb]{0,0,0}$b$}%
}}}}
\end{picture}%

   \caption{About {the} $R_2$ rule}
   \label{R2}
\end{figure}

Let us discuss this rule using Figure \ref{R2}. 
On the right, $a$ and $b$ signals are clearly not ordered ($a \bowtie b$ holds). But on the left,
which represents $a \vartriangleleft b$, one cannot determine whether a causal order holds 
($a \rightarrow b$) or whether it is a special case of absence of order $(a \| b)$.

The purpose of  {the} $R_2$ rule is to force one, when implementing  $(a \| b)$, to make  the case $a \vartriangleleft b$ have the same consequences as $b \vartriangleleft a$ or $a \bowtie b$.  {The} $R_2$ rule may then be rewritten
$$
\forall s, t \in \mathbf{t}R, (a\in \mathbf{i}R \wedge b \in \mathbf{o}R)  \vee  (a \in \mathbf{o}R \wedge b\in \mathbf{i}R) \; ; $$
$$\begin{array}{rcl}
\Big[ \; (\; s \vartriangleleft a \vartriangleleft  b \in \mathbf{t}R &\wedge& s  \vartriangleleft  b \vartriangleleft  a \in \mathbf{t}R\; ) \\
&\vee& s  \vartriangleleft (a \bowtie b) \in \mathbf{t}R \; \Big]  \\
 &\Rightarrow&   \\
\Big[ \; (\; s \vartriangleleft a \vartriangleleft b \vartriangleleft t \in \mathbf{t}R
&\vee&
s \vartriangleleft (a \bowtie b) \vartriangleleft t \in \mathbf{t}R \; ) \\
&\Rightarrow& s \vartriangleleft (a \| b) \vartriangleleft t \in \mathbf{t}R\;  \Big] 
\end{array}$$

\subsection{{The} $R_{2}^{\prime}$ Rule}
		
\begin{figure}[htp] 
   \centering

\begin{picture}(0,0)%
\includegraphics{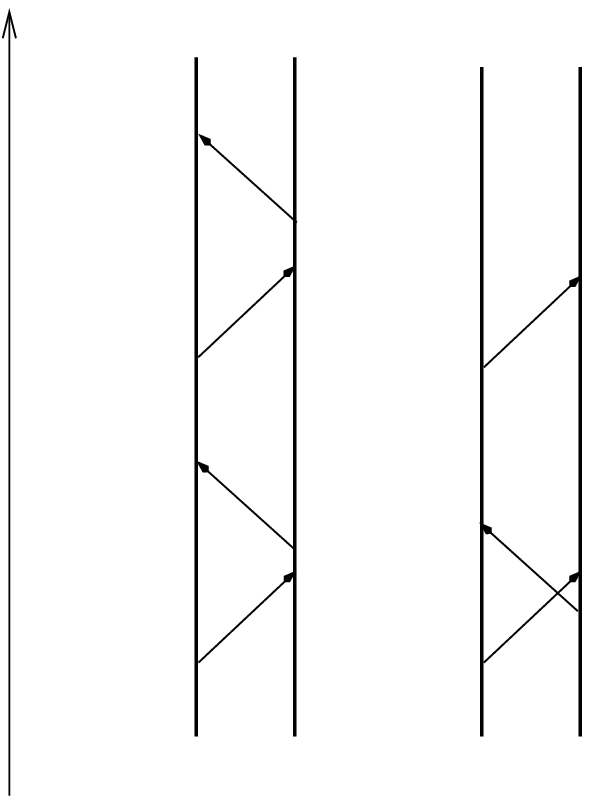}%
\end{picture}%
\setlength{\unitlength}{4144sp}%
\begingroup\makeatletter\ifx\SetFigFont\undefined%
\gdef\SetFigFont#1#2#3#4#5{%
  \reset@font\fontsize{#1}{#2pt}%
  \fontfamily{#3}\fontseries{#4}\fontshape{#5}%
  \selectfont}%
\fi\endgroup%
\begin{picture}(2674,3624)(859,-3448)
\put(1666,-3346){\makebox(0,0)[lb]{\smash{{\SetFigFont{10}{12.0}{\rmdefault}{\mddefault}{\updefault}{\color[rgb]{0,0,0}$E$}%
}}}}
\put(2116,-3346){\makebox(0,0)[lb]{\smash{{\SetFigFont{10}{12.0}{\rmdefault}{\mddefault}{\updefault}{\color[rgb]{0,0,0}$R$}%
}}}}
\put(1891,-2896){\makebox(0,0)[lb]{\smash{{\SetFigFont{10}{12.0}{\rmdefault}{\mddefault}{\updefault}{\color[rgb]{0,0,0}$a$}%
}}}}
\put(2971,-3346){\makebox(0,0)[lb]{\smash{{\SetFigFont{10}{12.0}{\rmdefault}{\mddefault}{\updefault}{\color[rgb]{0,0,0}$E$}%
}}}}
\put(3421,-3346){\makebox(0,0)[lb]{\smash{{\SetFigFont{10}{12.0}{\rmdefault}{\mddefault}{\updefault}{\color[rgb]{0,0,0}$R$}%
}}}}
\put(991,-16){\makebox(0,0)[lb]{\smash{{\SetFigFont{10}{12.0}{\rmdefault}{\mddefault}{\updefault}{\color[rgb]{0,0,0}$t$}%
}}}}
\put(1891,-1996){\makebox(0,0)[lb]{\smash{{\SetFigFont{10}{12.0}{\rmdefault}{\mddefault}{\updefault}{\color[rgb]{0,0,0}$b$}%
}}}}
\put(3151,-1591){\makebox(0,0)[lb]{\smash{{\SetFigFont{10}{12.0}{\rmdefault}{\mddefault}{\updefault}{\color[rgb]{0,0,0}$c$}%
}}}}
\put(3151,-2221){\makebox(0,0)[lb]{\smash{{\SetFigFont{10}{12.0}{\rmdefault}{\mddefault}{\updefault}{\color[rgb]{0,0,0}$b$}%
}}}}
\put(3151,-2941){\makebox(0,0)[lb]{\smash{{\SetFigFont{10}{12.0}{\rmdefault}{\mddefault}{\updefault}{\color[rgb]{0,0,0}$a$}%
}}}}
\put(1846,-1501){\makebox(0,0)[lb]{\smash{{\SetFigFont{10}{12.0}{\rmdefault}{\mddefault}{\updefault}{\color[rgb]{0,0,0}$c$}%
}}}}
\put(1846,-466){\makebox(0,0)[lb]{\smash{{\SetFigFont{10}{12.0}{\rmdefault}{\mddefault}{\updefault}{\color[rgb]{0,0,0}$d$}%
}}}}
\end{picture}%

   \caption{About {the} $R_{2}^{\prime}$ rule}
   \label{Rp2}
\end{figure}

{The} $R_2$ rule, under the form just discussed, may be too constraining in some cases. 
This is better seen again using Figure \ref{R2}.
 Let us assume that the two drawings represent the same non-ordered case $s \vartriangleleft (a \| b)$. 
On the left,  $R$ 
can be seen to have more information than $E$ on the time ordering of $a$ and $b$.
{$R$} also  knows that $E$ has to see the same order. On the other hand, $E$ cannot infer
 the time order seen by $R$, as it has no knowledge of the propagation delays.
 For instance, $E$ could infer $(a \bowtie b)$.
In that case, only $R$ can be authorized to make a decision. 
For instance,  $R$ can emit $d$ if $a$ and $b$ are not causally ordered while
 $a \vartriangleleft b$ holds (see Figure \ref{Rp2}).

If an event $c$ is emitted with the same type as $a$,  $c$ must exist whatever the order seen by $R$, because $c$ is emitted by $E$, which has less information than $R$. 
This is ensured by  {the} $R_{2}^{\prime}$ rule, which classically reads
$$\begin{array}{c}
\forall s,t \in \mathbf{t}R, \; (a, c\in \mathbf{i}R \wedge b\in \mathbf{o}R) \\
 \vee \quad (a, c\in \mathbf{o}R \wedge b\in \mathbf{i}R) \; ; \\
 (s a b t c\in \mathbf {t}R \wedge s  b a t\in \mathbf{t}R) \; \Rightarrow \;  s b a t c \in \mathbf{t}R
\end{array}$$
$R_2^\prime$ may then be rewritten
$$\begin{array}{c}
\forall s,t \in \mathbf{t}R,  (a, c\in \mathbf{i}R \wedge b\in \mathbf{o}R) \\
 \vee \quad (a, c\in \mathbf{o}R \wedge b\in \mathbf{i}R) \; ; \\
\Big[ \;  ( \; s \vartriangleleft a \vartriangleleft  b \vartriangleleft t \vartriangleleft c \in \mathbf{t}R \; \wedge \; s  \vartriangleleft  b \vartriangleleft  a \vartriangleleft t \in \mathbf{t}R \; ) \\
\vee  \quad s  \vartriangleleft (a \bowtie b) \vartriangleleft t \vartriangleleft c  \in   \mathbf{t}R \; \Big] \\
\Rightarrow \quad  s \vartriangleleft (a \| b) \vartriangleleft t \vartriangleleft c   \in  \mathbf{t}R
\end{array}$$

\subsection{{The} $R_{3}$ Rules}
{The} $R_3$ rules concern the possibility (or impossibility) for two symbols to be mutually exclusive.
For instance, the  design of a logic device may result in the following property.
If $a$ and $b$ are respectively input and  output symbols,
sharing no causal order ($a \| b$), a reception of $a$  prevents $b$ from being emitted 
as soon as $a$ precedes $b$ ($a \vartriangleleft b$). 
If the design has to avoid such a possibility, the latter must be excluded by a rule. Classically, the corresponding rule reads
\begin{equation}
s a \in \mathbf{t}R \; \land \; s   b \in \mathbf{t}R  \quad \Rightarrow \quad s   a b \in \mathbf{t}R \label{truc3}
\end{equation}
Extended to any combination of symbols of the same type, this rule can be used  to classify 
logic devices:
\begin{itemize}
\item Two input symbols can be mutually exclusive  if the environment has to make a choice {(data communication devices)}.
\item Two output symbols can be mutually exclusive if the component has to make a choice (arbitration devices).
\end{itemize}
{The} $R_{3}$ rules aim at forbidding such choices. For instance, {data communication devices}
 are excluded with the following rule:
$\forall s \in \mathbf{t}R,  \; a \in \mathbf{i}R  \wedge  b \in \mathbf{i}R$  ; then (\ref{truc3}).\\
In a relativistic framework, the second part of expression (\ref{truc3}) must be replaced by
the following
$s \vartriangleleft  (a \| b) \in \mathbf{t}R$ \\
so that expression (\ref{truc3}) should be rewritten
\begin{eqnarray}
\Big[  s \vartriangleleft (a \bowtie b) \in \mathbf{t}R &\vee& ( s \vartriangleleft  a \in \mathbf{t}R \; \land \; s \vartriangleleft  b \in \mathbf{t}R  )  \Big] \nonumber \\
&\Rightarrow& s \vartriangleleft  (a \| b) \in \mathbf{t}R \label{rel3} 
\end{eqnarray}
Inserting conditions bearing on different types of symbols before expression (\ref{rel3})
generates a group of three different $R_3$ rules.

\textbf{$R^\prime_{3}$ rule}: this first rule defines the most constrained class of components, where
 neither the environment, nor the component can make a choice (Muller-C is in this class)
$$\forall s \in \mathbf{t}R, \quad a \neq b \in{\bf a}R \; ; \quad \mbox{then (\ref{rel3}).}$$
		
\textbf{$R^{\prime\prime}_{3}$ rule}: this second rule accepts {data communication devices}, 
as the corresponding condition
does not constrain two inputs (a choice can be made on two inputs only)
$$\forall s \in \mathbf{t}R, \quad a \neq b \in{\bf a}R \quad a \notin \mathbf{i}R \vee b \notin \mathbf{i}R \; ; \quad \mbox{then (\ref{rel3}).}$$		

\textbf{$R^{\prime\prime\prime}_{3}$ rule}: according to this third rule, not only two inputs but also two outputs can be mutually exclusive (arbiters are in this class)
$$\forall s \in \mathbf{t}R, (a \in \mathbf{i}R \wedge b \in \mathbf{o}R) \vee (a \in \mathbf{o}R \wedge b \in \mathbf{i}R)  \; ; \mbox{then (\ref{rel3}).}$$		

The three $R_3$ rules have been ordered according to decreasing constraints put on specifications.
They allow one to generalize  Udding's classes of DI-components to the relativistic case.

\subsubsection{Relativistic DI-components}

By definition, relativistic DI-components  
have specifications that obey  $R_{0}$, $R_{1}$, $R^\prime_{2}$ and $R^{\prime\prime\prime}_{3}$
rules.

\section{Classical rules}

We now discuss how relativistic traces reduce to classical traces when communications 
can be considered to occur in a classical spacetime ({i.e.} assuming that 
propagations and time delays are totally controlled, at the level of elementary
logical operations). Accordingly, relativistic rules reduce to their classical analogs.  
This reduction amounts to a mapping that transforms R-traces into classical traces.
In particular, the set of symbols $a$, $b$, $\vartriangleleft$, $\bowtie$ is reduced when
 operations are restricted to a classical environment which is more {constraining}, as symbol $\bowtie$ is not supported any more.

Let us note that an R-trace {describes} a partial time-ordering that applies in particular to
\textit{propagations}, although a classical trace describes a total time-ordering which refers to
 \textit{points} in spacetime. Both notions however coincide when applied to events,
{i.e.} to signals at the input or output ports of a component. Hence, although a symbol $a$ does not a priori denote the same object in a R-trace as in a trace, it is legitimate to define its mapping
from a R-trace to a classical trace. One obtains the following  mapping.
\begin{itemize}
\item $a \vartriangleleft b \in \mathbf{t}R$ is mapped onto $ab  \in \mathbf{t}R$.
\item $(a \bowtie b)  \in \mathbf{t}R$ is mapped onto $ab \in \mathbf{t}R \; \wedge \;  ba \in \mathbf{t}R$.
\item $a \| b \in \mathbf{t}R$ is mapped onto $ab \in \mathbf{t}R \; \wedge \;  ba \in \mathbf{t}R$.
\end{itemize}
This mapping amounts to replacing the absence of time-order, unavoidable in a relativistic framework,
by the two possible orders, as imposed by the syntax of classical traces.
As illustrated by spacetime diagrams, one R-trace maps onto a pair of classical traces.
This corresponds to  the two different geometrical situations that can result when  two intersecting slanted arrows are made to separately intersect the two vertical lines representing a component and its environment. Both resulting traces have to be present in the specification of the component.
Indeed, all relativistic rules end with the form
$$\Rightarrow (a \| b) \in \mathbf{t}R$$
which is mapped onto the expression
$$  \Rightarrow \; ab \in \mathbf{t}R \; \wedge \;  ba \in \mathbf{t}R  $$
In other words, all these rules concern the absence of causal order and their restriction to classical spacetime amounts to imposing the presence of the two opposite orders.
The resulting rules are easily seen to correspond  to the standard ones.
In fact, one can note that {it is} {precisely} because Udding's rules have the property of imposing the presence of the two opposite orders that they can be obtained as  restrictions of  relativistic rules.
 This is the reason why the notion of DI-systems can be generalized to relativistic spacetime with its partial time-ordering.

\section{Conclusion}

The relativistic and classical frameworks lead to conceptions of time that differ in a significant way.
Time-simultaneity and total ordering in time are classical properties. 
{In contrast, relativistic time can only satisfy a partial ordering  to remain compatible with observer-dependent
propagation delays.}
Practical methods {depending on the classical properties of time} may reach a point where they cease to be {sufficient for a circuit to properly function, due to propagation delays}. The models used to design logic devices can
then be improved by {releasing these properties and looking for} a better implementation of relativistic causality. 
In particular, causal relations {can be made} compatible with
the impossibility for some events to satisfy a time order \cite{matherat2003}.

{Delay Insensitivity represents an efficient criterion for designing functional clockless circuits
in a simple way. But usual models}, such as the trace formalism, rely on a classical representation
of time-ordering and causality. {Nonetheless, traces can be generalized to  R-traces which 
provide a similar description of the dialog between a DI component and its environment
 while being compatible with  relativistic causality.}
The usual formalism \cite{udding-these,udding86,snepscheut85} is recovered  when 
 implementations of components are restricted to 
a classical environment.  

An R-trace structure describes the specification of a component-environment pair as the  list of all time orders that can hold between the communication events of  the pair dialog.
 When rewritten, Udding's rules keep a simple form that clarifies their relation  with causality properties.
{Although the R-trace formalism, when applied to
 specifications of DI components, takes a form similar to its classical analog \cite{Ebergen-thesis,ebergen91},} interpretations differ significantly.
R-trace structures describe the time-order and causality relations of all events, 
including {signal} propagations, while classical structures only refer to events that are localized in time.  

{The biggest merit of the R-trace formalism is to clarify the semantics underlying Udding's rules by exhibiting their
intrinsic relation with time-ordering and physical causality. 
Further theoretical insight, with potential consequences for applications, 
could be gained by comparing R-traces to  
other existing formalisms dealing with Delay Insensitivity. Criteria have indeed been 
developed to perform such comparisons as, for instance, the strength of semantic properties in equivalence
relationships \cite{vanGlabbeek:concur90}.    
}

{Besides a clear representation of causal relations, another merit of the  R-trace formalism
 is to replace the description of a distributed system as a set of 
individual components that communicate by a composition of  component-environment pairs.} 
This rather describes  the composition of \textit{interfaces}, where \textit{interface} means the dialog between the two members of a pair.
{In contrast to alternative descriptions of causality,
 such as Petri nets, this conception allows a simple 
composition of multiple components, by considering the latter as a multiplication of interfaces. This 
approach also opens} new and interesting possibilities for solving questions raised by the
 synchronization of complex distributed systems \cite{LamportACM78,Lamport1986,Mattern}.
{A further} advantage of the formalism is its ability to include in the description, 
with minor changes, components in motion. This could also allow the treatment of 
  distributed algorithms on a network of mobile computers within the formalism of DI systems.

\end{document}